\documentclass[10pt, a4paper]{article}

\usepackage[T1]{fontenc}
\usepackage[utf8]{inputenc}
\usepackage{calc}
\usepackage{indentfirst}
\usepackage{fancyhdr}
\usepackage{graphicx,epstopdf}
\usepackage{lastpage}
\usepackage{ifthen}
\usepackage{float}
\usepackage{amsmath}
\usepackage{amssymb}
\usepackage[right]{lineno}
\usepackage{setspace}
\usepackage{enumitem}
\usepackage{mathpazo}
\usepackage{booktabs} 
\usepackage{titlesec}
\usepackage[dvipsnames]{xcolor}
\usepackage{soul}
\usepackage{multirow}
\usepackage{microtype}
\usepackage{attrib}
\usepackage{upgreek}
\usepackage{array}
\usepackage{tabularx}
\usepackage{pbox} 
\usepackage{ragged2e} 
\usepackage[margin=25mm]{geometry}
\usepackage{marginfix} 
\usepackage{xstring}
\usepackage[backend=biber, style=numeric]{biblatex}
\newcommand\blfootnote[1]{%
  \begingroup
  \renewcommand\thefootnote{}\footnote{#1}%
  \addtocounter{footnote}{-1}%
  \endgroup
}

\newcommand\blue[1]{{#1}}

\title{Bernoulli principle in ferroelectrics}

\author{Anna G. Razumnaya $^{1}$, Yuri Tikhonov $^{2}$, Dmitrii Naidenko $^{3}$,\\ Ekaterina Linnik $^{3}$ and Igor Lukyanchuk $^{2}$}
\date{}

\begin{document}

\maketitle

\begin{abstract}
Ferroelectric materials, characterized by spontaneous electric polarization, exhibit remarkable parallels with fluid dynamics, where polarization flux behaves similarly to fluid flow. Understanding polarization distribution in confined geometries at the nanoscale is crucial for both fundamental physics and technological applications.
Here, we show that the classical Bernoulli principle, which describes the conservation of the energy flux along velocity streamlines in a moving fluid, can be extended to the conservation of polarization flux in ferroelectric nanorods with varying cross-sectional areas.
Geometric constrictions lead to an increase in polarization, resembling fluid acceleration in a narrowing pipe, while expansions cause a decrease. Beyond a critical expansion, phase separation occurs, giving rise to topological polarization structures such as polarization bubbles, curls and Hopfions. This effect extends to soft ferroelectrics, including ferroelectric nematic liquid crystals, where polarization flux conservation governs the formation of complex mesoscale states.
\end{abstract}

\blfootnote{
$^{1}$ \,Jozef Stefan Institute, Condensed Matter Physics Department, Ljubljana, 1000, Slovenia;
$^{2}$ \, University of Picardie, Laboratory of Condensed Matter Physics, Amiens, 80039, France;
$^{3}$ \,  Southern Federal University, Rostov-on-Don, 344090, Russia;}

\section{Introduction}

The Bernoulli principle, named after Daniel Bernoulli, who first published the concept in his book \textit{Hydrodynamica} in 1738~\cite{Bernoulli1738}, is one of the most fundamental concepts in fluid dynamics, playing a pivotal role in understanding the behavior of fluids in motion. It describes the relationship between the pressure, velocity, and potential energy in a steady, incompressible flow of an ideal fluid through a tube of varying cross-sectional areas~\cite{Landau6,Batchelor2000}. Specifically, it asserts that a decrease in the tube cross-section leads to an increase in the fluid velocity and a decrease in pressure, and vice versa. This simple yet profound principle has been instrumental in a wide range of applications, from aerodynamics, and various engineering disciplines to medical applications related to blood flow, and serves as a cornerstone.

While traditionally applied to fluids, the Bernoulli principle can also be extended to flux-conserving systems. One such system is ferroelectric materials that exhibit spontaneous electric polarization,  which can be modulated by external electric fields, mechanical constraints, or temperature variations. The 21st century inaugurated a revolution in the physics of ferroelectrics~\cite{Das2020}, introducing a fundamentally new understanding of matter through topology.
Over the past two decades, both theoretical and experimental findings have demonstrated that ferroelectricity remains robust at the nanoscale. This breakthrough revealed their ability to support a diverse range of topological states, including stable domain walls, vortices, cylindrical and bubble domains, and labyrinthine structures in ferroelectric thin films and superlattices~\cite{Stephanovich2003,Kornev2004,Zubko2012,Yadav2016,Das2019,Stachiotti2011,Guo2022,Junquera2023,Wang2023,Lukyanchuk2024,Han2025}. A variety of topological polarization structures have been directly observed using advanced high-resolution techniques such as piezoresponse force microscopy (PFM), scanning transmission electron microscopy (STEM), and coherent X-ray 
diffraction~\cite{Gruverman2019,Cabral2023,Karpov2017}.
 \blue{Although topological polarization structures such as vortex domains and polarization bubbles have been predicted theoretically in ferroelectric nanorods~\cite{Naumov2004,Prosandeev2008,Lahoche2008,Liu2017,Lich2022,Kondovych2023} their direct experimental observation remains a challenge.
 }

In ferroelectrics, topological matter is governed by nonlocal electrostatic fields, meaning its properties are influenced not only by the material internal structure but also by the overall geometry of the system. Fundamentally, the tendency to minimize unfavorable depolarization electrostatic forces arising from bound charges induced by polarization, drives ferroelectrics to maintain a divergence-free polarization distribution~\cite{Lukyanchuk2024}. This topological constraint is similar to the divergence-free nature of the velocity field in fluids, which ultimately determines the topology of the vector field~\cite{Moffatt1992,Arnold2021}. As in fluid dynamics, this leads to the conservation of polarization flux along polarization streamlines, giving rise to the Bernoulli effect.  

A wide range of theoretical approaches has been successfully applied to study topological polarization states in ferroelectrics, including atomistic simulations that fully capture microscopic structures, as well as continuum methods, based on numerical solution of the Ginzburg-Landau (GL) equations, coupled with electric and elastic fields, which are effective at large scales. However, as systems become increasingly complex and multiscale, the simulations come with a high computational cost.
Our work introduces a physical perspective based on polarization flux conservation,  analogous to flux conservation in fluids. This analogy provides an intuitively unifying insight to interpret and predict many of the complex topological structures in ferroelectrics~\cite{Lukyanchuk2024}. 

We extend the Bernoulli principle to ferroelectric systems by examining the behavior of polarization flux in cylindrical ferroelectric pipes, specifically in ferroelectric nanorods with varying cross-sectional areas. 
Such ferroelectric nanorods with sub-10\,nm diameters have been experimentally realized using various bottom-up synthesis techniques, including hydrothermal and solvothermal methods~\cite{Varghese2013}. Stable and tunable ferroelectric behavior has been observed in these systems.

By drawing an analogy with classical hydrodynamics, we derive a generalized Bernoulli equation that accounts for the internal energy contributions of the polarization field, governed by the Ginzburg-Landau-Devonshire (GLD) functional. Our findings demonstrate that constrictions in ferroelectric nanorods lead to an increase in polarization, akin to the acceleration of a fluid in a narrowing pipe, whereas expansions result in a decrease in polarization. Beyond a critical expansion threshold, the system undergoes an instability, leading to the formation of ferroelectric closed-loop domain structures, polarization bubbles, curls, and Hopfions. These topologically complex states reveal novel aspects of polarization dynamics in nanostructured ferroelectric materials. 

Beyond this, our study expands to soft ferroelectrics, particularly recently discovered ferroelectric nematic liquid crystals~\cite{Nishikawa2017,Chen2020,Kumari2024}, where polarization flux conservation plays a crucial role in the formation and manipulation of topological states. The ability to control these structures using external electric and elastic fields, as well as temperature variations, opens new avenues for applications in nanoelectronics and energy storage. By establishing a direct connection between fluid dynamics and polarization flux in ferroelectric materials, our findings provide a fundamental framework for understanding and engineering novel functionalities in nanostructured ferroelectric systems.

The paper is organized as follows. In Section~2, we describe the theoretical framework and computational methods used in our study. We introduce the GLD approach, which provides a phenomenological description of ferroelectric polarization dynamics, followed by the phase-field modeling technique used to numerically investigate the system. The additional technical details on the phase-field simulations and material parameters are given in Appendix~A. Section~3 presents our main results. We begin with a discussion of the classical Bernoulli effect in fluids as a reference point. We then introduce the concept of the Bernoulli effect in ferroelectrics, demonstrating how polarization flux conservation governs polarization redistribution in varying geometries. Finally, we provide phase-field modeling results that confirm the emergence of topological polarization states, such as vortex domains, Hopfions, and polarization bubbles and curls, under specific materials and geometric constraints. In Section~4, we discuss the implications of our findings, highlighting the role of material anisotropy and possible experimental approaches for verifying the predicted effects. We also outline potential applications of the ferroelectric Bernoulli effect in nanoelectronics and soft ferroelectrics. Section~5 summarizes our key results.

\section{Materials and Methods}

\subsection{Ginzburg-Landau approach. }
\label{GLapproach}

The Ginzburg-Landau-Devonshire (GLD) theory forms the foundational framework for describing the polar states in ferroelectrics~\cite{Lines2001,Rabe2007,Strukov2012}. Central to this theory is the concept of an order parameter, which in the case of ferroelectrics is the spontaneous polarization $\mathbf{P}$. The GLD approach is based on minimizing the free energy functional, which accounts for the spatial variation of polarization and its interactions with elastic and electric fields, and is expressed as:
\begin{equation}
\mathcal{F} = 
\int_{}^{} ( F_{GL} + F_{grad} + F_{\varphi}+ F_{elast} )d^{3}r.
\label{Functional}
\end{equation} 
This functional includes the Ginzburg-Landau (GL) energy density of the uniform state $F_{GL}$, the gradient energy density $F_{grad}$ associated with the spatial variation of polarization, and the electrostatic $F_{\varphi}$ and elastic $F_{elast}$ contributions, which describe the coupling of ferroelectric polarization with electric and elastic fields, respectively. 

To explore the Bernoulli effect in ferroelectrics, we begin with the simplest case of a uniaxial ferroelectric, where the spontaneous polarization aligns with the crystal anisotropy axis. In this case, the order parameter reduces to a single scalar component of polarization, $P$, which can take either positive or negative values, depending on the orientation of the spontaneous polarization relative to the axis. The uniform GL energy density  $F_{GL}$ of a polar state is  expressed as a function of the order parameter, typically expanded up to 4th order in $P$ :
\begin{equation}
F_{GL} =  \alpha(T-T_c) P^2 + b P^4\,.
\label{FGL}
\end{equation}
Here, $T$ is the temperature, $T_c$ is the Curie temperature at which the phase transition occurs, and $\alpha$ and $b$ are positive, material-dependent coefficients. The coefficient $\alpha$ is related to the Curie constant $C$ via $\alpha = 1 /2 \varepsilon_0 C$, where $\varepsilon_0 = 8.85 \times 10^{-12}~\mathrm{C\,V^{-1}m^{-1}}$ is the vacuum permittivity.
Above $T_c$, the system is in a paraelectric phase, with no polarization, since $P = 0$ minimizes the free energy, reflecting a symmetric phase without spontaneous polarization. Below $T_c$, the coefficient $\alpha(T - T_c)$ in $F_{GL}$ becomes negative, and the term $bP^4$ ensures the stability of the free energy, preventing $P$ from diverging. The system minimizes its free energy for a non-zero polarization, revealing a continuous, second-order transition to the ferroelectric phase.
The energy profile of the uniform state, ${F}_{GL}(P)$, exhibits a characteristic double-well structure, with two degenerate minima corresponding to stable ferroelectric states of opposite polarization. These states are defined by the equilibrium spontaneous polarization $P_0 = \pm \left[\alpha(T_c - T)/2b\right]^{1/2}$, representing two energetically equivalent orientations along the anisotropy axis. In typical oxide ferroelectrics, the magnitude of $P_0$ is on the order of $0.1~\mathrm{C/m^2}$. In some materials, the ferroelectric phase transition occurs discontinuously as a first-order transition. This scenario is described by a negative coefficient $b$ and the inclusion of a stabilizing sixth-order term $cP^6$ in ${F}_{GL}$, with $c > 0$. The presence of this term ensures the boundedness of the free energy. The subsequent analysis does not qualitatively depend on the order of the phase transition.

The gradient term in functional (\ref{Functional}), 
\begin{equation}
F_{grad}=\gamma (\nabla P)^2,
\label{Fgrad}
\end{equation}
 with $\gamma>0$, accounts for the energy cost of spatial variations of polarization, in the polarization domain structures.

An important aspect in understanding the properties of the ferroelectric state is accounting for the coupling of polarization with electric and elastic fields. The energy effect of an electric field $\mathbf{E}=-\nabla \varphi$ (where $\varphi$ is electrostatic potential) is described by the electrostatic contribution 
\begin{equation}
{F}_{\varphi} =   \mathbf{P} \nabla\varphi-\frac{1}{2}\varepsilon_0\varepsilon_b(\nabla \varphi)^2\,,
\label{Fel}
\end{equation}
where the first term represents the coupling between the electric field and corresponds to the work done by the electric field in polarizing the material.  The second term is the Lagrange density of the electrostatic energy,  and $\varepsilon_b\sim 10$ is the dielectric constant characterizing the non-polar background of the material~\cite{Mokry2016}.
Variational minimization of the GLD functional (\ref{Functional}) with respect to the electrostatic potential $\varphi$ results in the core electrostatic relation  $\mathrm{div}(\mathbf{P}+\varepsilon_0\varepsilon_b\mathbf{E})=0 $, that demonstrate that the so-called depolarization electric filed $\mathbf{E}_{dep}$ is produced by the bound charges with density $\rho_b=-\mathrm{div}\mathbf{P}$, emerging due to polarization inhomogeneities. The electrostatic depolarization energy associated with this field in typical ferroelectrics is significantly large, exceeding the favorable for the formation of the ferroelectric state GL energy by one-–two orders of magnitude. Consequently, ferroelectrics tend to minimize the formation of depolarization fields and bound charges. 

A well-known manifestation of the depolarization field reduction is the formation of Landau-Kittel domains in ferroelectric slabs~\cite{Landau1935,Kittel1946,Bratkovsky2000}. At the nanoscale, in ultra-thin ferroelectric films and superlattices, these domains appear as periodic patterns of domains with soft polarization profile~\cite{Guerville2005}, having the structure of oppositely rotating polar vortices and bubble structures~\cite{Kornev2004,Zubko2012,Yadav2016,Das2019,Pavlenko2022b} arranged within the plane of the ferroelectric film. More broadly, the rich diversity of topological states observed in nanostructured ferroelectrics over the past decades is fundamentally driven by the minimization of the depolarizing field effect. 

\blue{
In typical ferroelectrics, the depolarization field energy density associated with surface-bound charges in a uniform polar state is estimated as
$F_{\mathrm{dep}} \approx {P_0^2}/{2\varepsilon_0 \varepsilon_b} \sim 10^7$–-$10^8~\mathrm{J/m^3}$.
This unfavorable electrostatic contribution far exceeds the favorable GL energy density,
$F_{\mathrm{GL}} \sim -{P_0^2}/{8\varepsilon_0 \chi} \sim -(10^5$
---$10^6)~\mathrm{J/m^3}$,
due to the large Curie--Weiss susceptibility in oxide ferroelectrics, $\chi \approx C/2T_c \sim 100$~\cite{Strukov2012}.} As a result, to mitigate the depolarizing electric field, the system seeks to reduce the number of bound charges responsible for its generation. In other words, the polarization field naturally tends to minimize charge-inducing divergence, thereby satisfying a nearly divergence-free condition ~\cite{Lukyanchuk2024},
\begin{equation}
    \mathrm{div}\mathbf{P}\approx 0.
    \label{divP=0}
\end{equation} 

\blue{This condition, which is the main factor for the emergence of topological structures of the polarization flux}, is remarkably analogous to the incompressibility condition $\mathrm{div}\,\mathbf{v}=0$ in hydrodynamics, governing fluid flow with velocity $\mathbf{v}$. In ferroelectrics, it ensures the conservation of polarization flux along polarization streamlines, leading to a distinctive Bernoulli-like effect in the elongated tube-like structures with varying cross-sections. This phenomenon is the focus of this article. Note that the constraint (\ref{divP=0}) significantly enhances the complexity of the Lagrange variational minimization of the functional (\ref{Functional}), resulting in a greater variety of possible scenarios.

Ferroelectric materials also exhibit significant sensitivity to elastic deformations, making it essential to incorporate strain-dependent terms into the free energy expression within the framework of the GLD theory. Strain in a solid is quantified by the strain tensor,
$u_{ij} = \frac{1}{2} \left(\partial_j  u_i  + \partial_i u_j \right) $
where $u_i$ is the displacement vector, representing the deviation of a material point from its position in the non-deformed state.
This strain tensor consists of six independent components, which are defined by the symmetry of the crystal. 
Here, we illustrate the elastic contribution considering only the uniform deformation of the crystal, described by $u_{ij}=u\delta_{ij}$, and present a more comprehensive treatment within the full phase-field approach. The isotropic part of the elastic energy density is written as:
\begin{equation}
F_{elast} =  -Q {P}^2(3u)+\frac{1}{2}K(3u)^2, 
\label{Felast}
\end{equation}
where the first term represents a coupling between elastic strain and polarization with a coupling electrostrictive constant $Q$, and the second term adheres to Hooke’s law, indicating that the elastic energy stored in a solid is quadratically dependent on distortion, with $K$ representing one of the elastic constants. Notably, in the case of a uniform or nearly uniform polar state, the elastic contribution can be incorporated into the GL energy by integrating the GLD functional over the strain field, which effectively leads to the renormalization of the coefficient $b$, such that 
$b \to b-Q^2/2K$~\cite{Strukov2012}. Then, the deformation is related to polarization as $u=(Q/K)P^2$. This renormalization allows the elasticity effects to be disregarded in certain model cases.

\subsection{Phase-Field Simulations}

The analytical approach to ferroelectrics presented in the previous section employs simplified models, such as the one-component ferroelectric model, to isolate key physical mechanisms. These include the minimization of depolarization effects, which enforces polarization flux conservation along streamlines and drives the emergence of topological states. As these fundamental principles are expected to hold universally across different ferroelectric materials, this approach remains particularly valuable for developing general theoretical predictions.

To move beyond the model cases and assess the robustness of our concepts under more realistic conditions, we employ phase-field modeling. Phase-field simulations provide a powerful tool for studying polarization dynamics in ferroelectrics, complementing the analytical methods discussed in the previous Section.
Unlike analytical models, which deliberately reduce complexity to highlight core physical effects, phase-field simulations introduce no simplifying assumptions about polarization distribution, electrostatic and elastic interactions, or material anisotropy. Instead, they are based on a direct minimization of the full GLD energy functional (\ref{Functional}), which includes Ginzburg-Landau, gradient, electrostatic, and elastic energy contributions. They correctly account for all relevant material parameters, including the complex functional structure and tensorial form of material parameters, and complex geometry of the system and appropriate boundary conditions. This approach allows us to study the system in a fully self-consistent manner. 

The detailed formulation of the complete GLD functional and numerical implementation of the phase-field simulations are provided in Appendix~A. By solving the coupled partial differential equations derived from this energy functional, phase-field modeling enables us to investigate the Bernoulli effect in realistic ferroelectric systems. These simulations yield several key insights. They confirm that the analytically derived divergence-free structure of the polarization field and the conservation of polarization flux along streamlines remain valid even in complex geometries with varying cross-sections. They also reveal an instability that triggers the spontaneous emergence of polarization states with non-trivial topology when certain geometric and material parameters exceed a critical threshold, thereby validating analytical predictions of domain nucleation instability. Furthermore, they uncover additional topological features such as curled and Hopfion-like polarization structures, and polarization bubble domains, which are not considered by simplified analytical models but emerge naturally through numerical energy minimization.

\section{Results}


\begin{figure*}
\centering
\includegraphics[width=\linewidth]{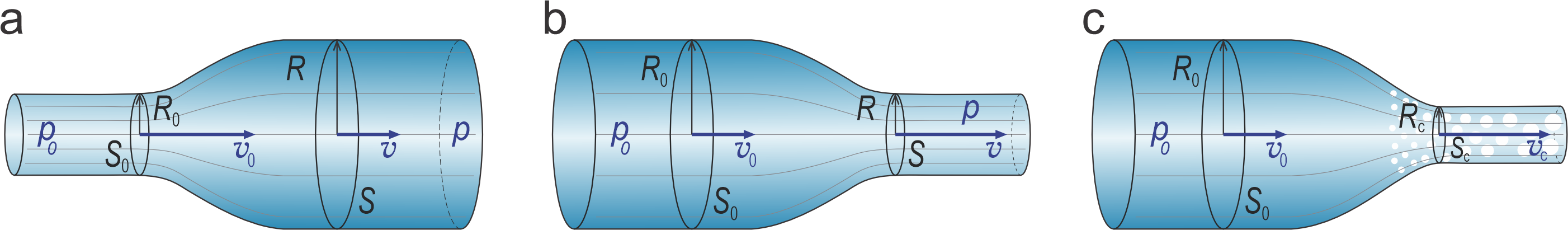}
\caption{\textbf{Bernoulli effect and cavitation in fluids}. 
(a) Flow through an expanding section leads to a decrease in velocity $v$ and an increase in pressure $p$.  
(b) Constriction causes an increase in $v$ and a decrease in $p$.  
(c) Cavitation occurs with further constriction of the tube below critical radius $R_c$, when pressure drops below the vapor pressure, leading to bubble formation.   
}
\label{Fig1}
\end{figure*}

\subsection{Bernoulli Effect in Fluids}

The Bernoulli principle is a fundamental concept in fluid dynamics that describes the behavior of a moving fluid. It is derived from the conservation of energy and plays a key role in explaining the interrelation between velocity, pressure, and potential energy in a steady, incompressible flow of an ideal fluid~\cite{Landau6,Batchelor2000}.
For a steady, incompressible, and inviscid (zero-viscosity) horizontal flow, Bernoulli principle states that the total mechanical energy, consisting of pressure energy and kinetic energy, remains constant along a flow streamlines:
\begin{equation} p + \frac{1}{2} \rho v^2 = \text{const},
\label{EqBernoulliFluid} \end{equation}
where $p$ is the static pressure, $\rho$ is the fluid density, and $v$ is the flow velocity. The kinetic energy density, also known as the dynamic Bernoulli pressure $p_B = \frac{1}{2} \rho v^2$, allows the Bernoulli equation to be expressed in terms of the total pressure of the fluid, $p_{\text{tot}} = p + p_B$, which remains constant along a streamline.

To explore the Bernoulli effect, consider a fluid flowing with an initial velocity $v_0$ through a tube of radius $R_0$ and cross-sectional area $S_0 = \pi R_0^2$. At a certain point, the tube changes its radius $R$, leading to either increasing, for $R>R_0$, or decreasing, for $R<R_0$, of the cross-sectional area $S = \pi R^2$ as illustrated in Fig.\,\ref{Fig1}a and Fig.\,\ref{Fig1}b respectively.
 To conserve mass, the fluid velocity $v$ must adjust in response to the change in geometry, following the continuity equation:  
\begin{equation}
\pi R_0^2 v_0 = \pi R^2 v.
\label{EqContinuityFluid}
\end{equation}  
This equation is a direct consequence of the general continuity condition $\mathrm{div}\,\mathbf{v} = 0$ for an incompressible fluid with constant density. Consequently, as the cross-section of the tube varies, the velocity changes according to  
\begin{equation}
v =  \frac{R_0^2}{R^2}  v_0 .
\end{equation}  

Now, by applying Bernoulli equation at two points, one at the initial section of the tube and the other at the modified section, we derive the relationship between pressure and velocity. 
Rearranging Bernoulli equation gives the following relationship between the static pressure at the modified and initial sections of the tube, 
\begin{equation}
p= p_0+\frac{1}{2} \rho (v_0^2 - v^2) = p_0+\frac{1}{2} \rho v_0^2 \left(1-\frac{R_0^2}{R^2} \right).
\label{EqPressure} 
\end{equation}
Notably, the last term on the right-hand side of Eq.\,(\ref{EqPressure}) is positive for $R > R_0$ and negative for $R < R_0$, highlighting a key manifestation of the Bernoulli effect: the fluid slows down in the wider section and accelerates in the narrower section. Slower fluid flow corresponds to higher pressure, whereas accelerated flow leads to lower pressure.

 Importantly, at the narrowing of the tube, below the critical radius $R_c = R_0 ( 1+{2p_0}/{\rho v_0^2}  )^{-1/2}$, the whole expression in Eq.\,(\ref{EqPressure}) becomes negative, making $p < 0$, which indicates instability of the flow.
This instability gives rise to a phenomenon known as cavitation. Cavitation is the process in which multiple vapor bubbles or cavities form in a fluid when the pressure drops below the vapor pressure $p_c$, as illustrated in Fig.\,\ref{Fig1}c. In engineering and technology, cavitation is typically considered a destructive effect, commonly occurring in pumps, propellers, and turbines, where rapid fluid motion and pressure variations take place. The local boiling of the fluid induced by cavitation leads to the stochastic formation and collapse of bubbles, generating shockwaves that can cause significant damage to surrounding surfaces.

\subsection{Bernoulli Effect in Ferroelectrics}

\begin{figure*} [b!]
\centering
\includegraphics[width=\linewidth]{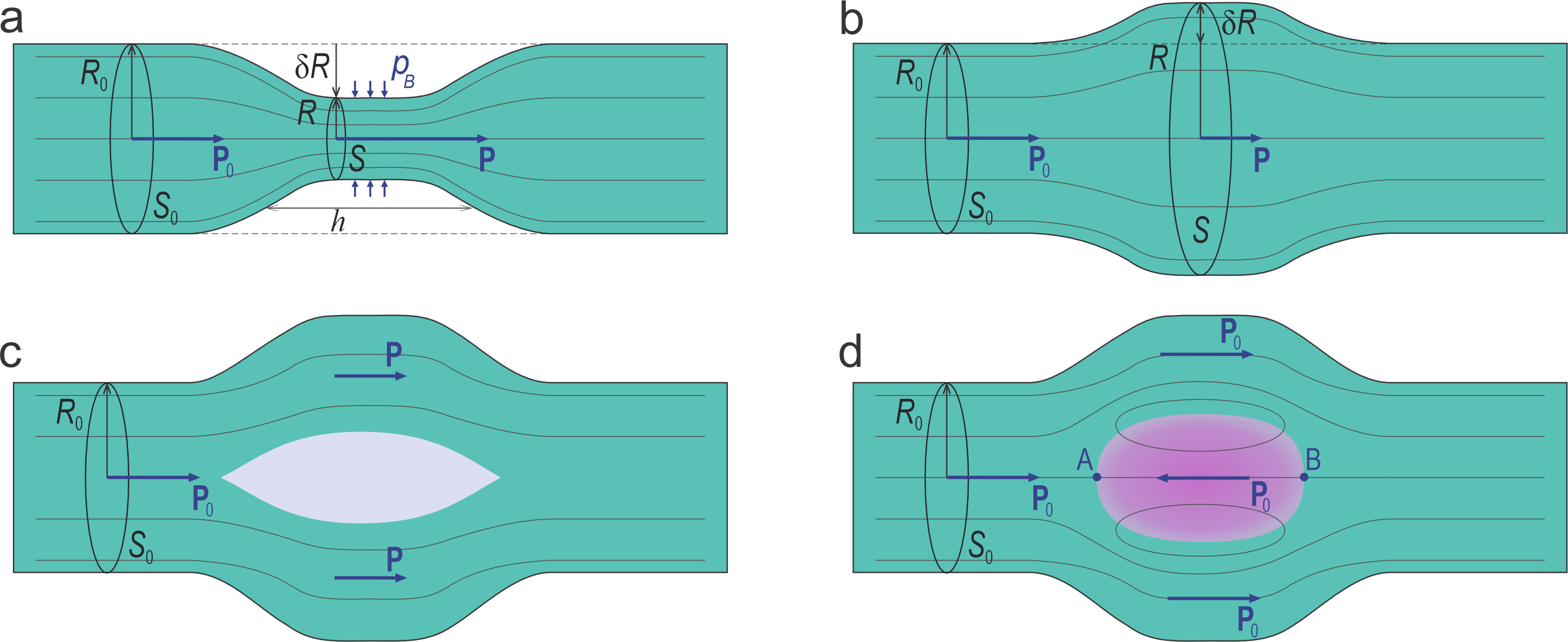}
\caption{
\textbf{Bernoulli effect and domain formation in ferroelectric fluxes}.   
(a) Flow through a constricted segment of the tube (shown in turquoise) results in an increase in polarization. A pressure $p_B$ is applied to decrease the tube radius within this segment.
(b) Flow through an expanded segment results in a decrease in polarization.
(c) Beyond a critical expansion, phase separation occurs, resulting in the formation of a void filled with a paraelectric phase (shown in gray) within the expanded segment, which minimizes the system's energy.
(d)  Formation of a counter-polarized domain (shown in violet) within the cavity, featuring a stable vortex-like configuration of streamlines, further reduces the system's energy. A and B are the singular points where the propagating polarization flux meets the counter-polarized domain.
}
\label{Fig2}
\end{figure*}

In this Section, we introduce the Bernoulli effect in ferroelectrics, drawing an analogy with fluid dynamics. This effect originates from the nearly divergence-free behavior of the polarization field, expressed as
$ \mathrm{div}\,\mathbf{P} \approx 0 $, which, as highlighted in Section~\ref{GLapproach}, plays a fundamental role in governing the polarization dynamics in ferroelectric nanostructures. To illustrate this concept, we consider an infinite cylindrical tube of radius $ R_0 $ that sustains a uniform ferroelectric polarization flux. To provide conceptual understanding, we begin with the simplest model, described in Section~\ref{GLapproach}: a uniaxial ferroelectric with polarization aligned along the tube axis.
 Within the tube, there exists a finite length segment of length $ l $ where the radius is modified to $ R $, either decreasing or increasing, as illustrated in Fig.\,\ref{Fig2}a and Fig.\,\ref{Fig2}b, respectively. Because of the divergence-free nature of the polarization field, the polarization flux remains conserved throughout the cylinder.
This conservation leads to an equation analogous to (\ref{EqContinuityFluid}):
\begin{equation}
    \pi R_0^2 P_0 = \pi R^2 P,
    \label{EqFlucConsrev}
\end{equation}
which relates the polarization $P_0$ in the infinite region to $P$ in the reshaped segment. 
In the infinite section of the cylinder, the polarization $P_0$ satisfies the equilibrium condition determined by the uniform GL equation, ${\partial {F}_{GL}}/{\partial P} = 0$, and the corresponding energy density is given by $F_0 = F_{GL}(P_0)$. However, in the reshaped segment, the polarization evolves according to $P = (R_0 / R)^2 P_0$, resulting in flux expansion when $R > R_0$ and flux compression when $R < R_0$. In both cases, the energy density $F_{\mathrm{GL}}(P)$ increases and exceeds its initial value $F_0 = F_{\mathrm{GL}}(P_0)$.

Remarkably, the concept of flux compression also arises in plasma physics, where magnetic fields are intensified through geometrical confinement. This principle, introduced by Andrei Sakharov in the 1960s\,\cite{Sakharov1966}, is employed in explosive magnetic flux compression generators to produce ultra-strong magnetic fields. Although based on a somewhat different mechanism than polarization flux compression, electromagnetic induction rather than electrostatic equilibrium, the analogy is striking. Both phenomena highlight the broad applicability of flux-conserving principles across diverse domains of physics.

We now examine the modification of the total GL energy in the polar phase, given by $\mathcal{F}_{GL} = V F_{GL}$, within a volume segment $V = \pi R^2 l$, under compression (see Fig.\,\ref{Fig2}a), where the segment radius decreases from $R_0$ to $R < R_0$. The case of expansion, where $R > R_0$ (Fig.\,\ref{Fig2}b), is treated analogously. The variation of the GL energy is given by  
 \begin{equation}
    \delta \mathcal{F}_{GL} = \delta (V{F}_{GL})=
    {F}_{GL}\delta V+V\delta {F}_{GL} 
    =\left[ {F}_{GL}  
    +  \frac{1}{2} R \frac{\partial {F}_{GL}}{\partial P}
    \frac{\partial P}{\partial R} \right] 2 \pi R l \delta R
    =\left[ {F}_{GL}  
   - \frac{\partial {F}_{GL}}{\partial P} P \right] 2 \pi R l \delta R
    \,,
    \label{EqBernoulliDerivation}
\end{equation}
where $\delta R=R-R_0<0$. In the last term, we have applied the condition $\partial P/\partial R = -2P/R$, which follows from flux conservation. The variation of the elastic part of the energy is not considered here.  
Assuming that the compression of the polarization flux is induced by a pressure $p_B$ acting on the interface of the ferroelectric phase, and equating the variation of the GL energy in Eq.~(\ref{EqBernoulliDerivation}) to the work done by this pressure, $\delta \mathcal{W} = - p_B 2 \pi R l \delta R$, we obtain  
\begin{equation}
   p_B = \frac{\partial F_{GL}}{\partial P} P - F_{GL}\,.
    \label{EqBernoulli}
\end{equation}
The pressure $p_B$ represents the Bernoulli\blue{-like} pressure of the polarization flux, analogous to the dynamic Bernoulli pressure in fluids, $p_B = \frac{1}{2} \rho v^2$. The latter can be derived from Eq.~(\ref{EqBernoulli}) by considering the kinetic energy density $F_{fluid} = \frac{1}{2} \rho v^2$ instead of $F_{GL}$ and varying it with respect to the velocity $v$. In the case of ferroelectrics, described by the GL energy (\ref{FGL}), the Bernoulli pressure~(\ref{EqBernoulli}) takes a specific form:
\begin{equation} 
    p_{B} =  \alpha(T-T_c) P^2 + 3 b P^4. 
\label{EqBernoulliGL} 
\end{equation}

Phase separation in ferroelectrics, analogous to cavitation in fluids, occurs when the Bernoulli pressure of the polarization flux becomes negative, $p_B < 0$. This happens when the condition $F_{GL} > (\partial F_{GL} / \partial P) P$ is met. Taking into account the continuity condition~(\ref{EqFlucConsrev}), we find that instability occurs when the radius of the expanded segment exceeds a critical value, $R_c$, at which the polarization minimizes to its critical amplitude, $P_c$. These threshold values are given by:  
\begin{equation}
   R_c = (3/2)^{1/4} R_0 \approx 1.1 R_0\,, 
   \quad 
   P_c = (2/3)^{1/2} P_0 \approx 0.82 P_0\,.
    \label{EqRc}
\end{equation}

Notably, in ferroelectrics, the instability with respect to phase separation occurs when the tube segment expands rather than contracts, as in the case of cavitation in fluids. This difference arises from the different functional dependencies of the energy functionals.  In ferroelectrics, the system's energy decreases with the expansion of the reshaped tube segment, whereas in fluids, energy decreases upon its contraction.

The Bernoulli-like pressure and phase-separation instability in ferroelectrics can be understood through the following considerations. In an infinitely extended section of the tube, the polarization is equal to its equilibrium value $P_0$, and the energy density $F_0 = F_{GL}(P_0) < 0$ corresponds to the minimum of $F_{GL}(P)$. When the polarization flux enters a reshaped segment, the energy density $F_{GL}(P)$ increases above $F_0$, regardless of whether the tube is contracting, causing a decrease in polarization $P$ relative to $P_0$, or expanding, leading to an increase in polarization $P$ relative to $P_0$.
 However, the total energy of the segment volume $V$, given by  
$
\mathcal{F}_{GL} = V F_{GL}=\pi R^2 l\, F_{GL},
$  
exhibits a different behavior with the variation of $R$ due to the contribution of the volume factor $\pi R^2 l$. Under compression, the total energy increases because $F_{GL}$ is negative, whereas during expansion, it initially decreases. The tendency of the flux entering the wide segment to expand in order to lower its total energy corresponds to a positive Bernoulli pressure $p_B$, similar to a gas expanding into free space due to its internal pressure.

A direct minimization of $\mathcal{F}_{GL}(P)$, where the energy density $F_{GL}(P)$ is defined by the functional (\ref{FGL}) and the polarization profile $P(R)$ follows from Eq.\,(\ref{EqFlucConsrev}), reveals that the total energy decreases with increasing volume until it reaches a minimum  
$
\mathcal{F}_{\min} = \pi R_c^2 l \,F_{GL} (P_c)
$  
at the critical radius $R_c$, as given by Eq.\,(\ref{EqRc}). Beyond $R_c$, further flux expansion raises its total energy above the minimum $\mathcal{F}_{\min}$, generating contraction forces that drive the flux back toward its lowest energy state at $R_c$. This regime change is marked by a sign change in the pressure $p_B$ from positive to negative once $R_c$ is exceeded. To maintain minimal energy, the polarization flux ceases to expand with the segment, stabilizing the cross-sectional area at $S_c = \pi R_c^2$, which leads to phase-separation instability. The stabilization of cross-sectional area occurs, for example, through the formation of a void cavity filled with the paraelectric phase, as illustrated in Fig.\,\ref{Fig2}c. The paraelectric phase has an energy $\mathcal{F}_{\text{para}} = \mathcal{F}_{{GL}}(0) = 0$. For $R > R_c$, the total energy of the system, where phase separation occurs between the ferroelectric and paraelectric phases, is given by  
$
\mathcal{F}_{\text{min}} + \mathcal{F}_{\text{para}} = \mathcal{F}_{\text{min}}.
$  
This energy is lower than that of a uniformly distributed ferroelectric flux across the segment cross-section, confirming the energetic favorability of phase separation.

An even more favorable configuration arises when this cavity is filled with a polarization domain oriented in the opposite direction, as shown in Fig.\,\ref{Fig2}d. In this arrangement, the polarization magnitude in both the propagating flux and the oppositely directed domain remains close to the equilibrium value \( \pm P_0 \). This allows the system to reach its minimum possible energy, with only a slight perturbation due to the energy of the domain wall separating oppositely oriented polarization states.  
Points A and B, where the propagating flux meets the polarization of an oppositely oriented domain, are the singular points of the polarization vector fields. Despite the head-to-head and tail-to-tail orientations of the polarization vectors approaching these points from opposite sides, no bound charge is formed. This is because the colliding fluxes are deflected sideways, generating a saddle-like configuration of the polarization streamlines and ensuring the divergence-free nature of the vector field. Such points are analogous to stagnation points in fluid flow in hydrodynamics. Overall, the total polarization flux remains conserved along the entire length of the tube, as the polarization field lines in the expanded segment form a closed, divergence-free vortex structure along the domain walls, resulting in zero net contribution to the total flux.
In the next chapter, we explore cases where the polarization vector can rotate in space, giving rise to a more intricate emerging counter-oriented textures with twisted structure.

\subsection{Modeling of the Bernoulli Effect in Ferroelectrics}

We examine now the emergence of the Bernoulli effect in characteristic physical ferroelectric systems, minimizing the total GLD energy functional (\ref{Functional}) using the phase-field simulation technique, detailed in Appendix~A.  
We begin by exploring an example of ferroelectric materials with weak anisotropy. A notable case is solid Pb(Zr$_{1-x}$Ti$_x$)O$_3$ with $x\sim0.4$, near the morphotropic boundary, where the crystal lattice only weakly pins the polarization vector~\cite{Lukyanchuk2020}. Another relevant system is the so-called soft ferroelectrics, where polarized molecules can freely rotate and move. This is particularly relevant to the recently discovered ferroelectric nematic liquid crystals~\cite{Nishikawa2017,Chen2020,Kumari2024}. In soft ferroelectrics, which exhibit both micro- and macroscopic topological states, the results must be appropriately rescaled from the nanoscale to the micro- and macroscales. 

\begin{figure*}[p]
\centering
\includegraphics[width=\linewidth]{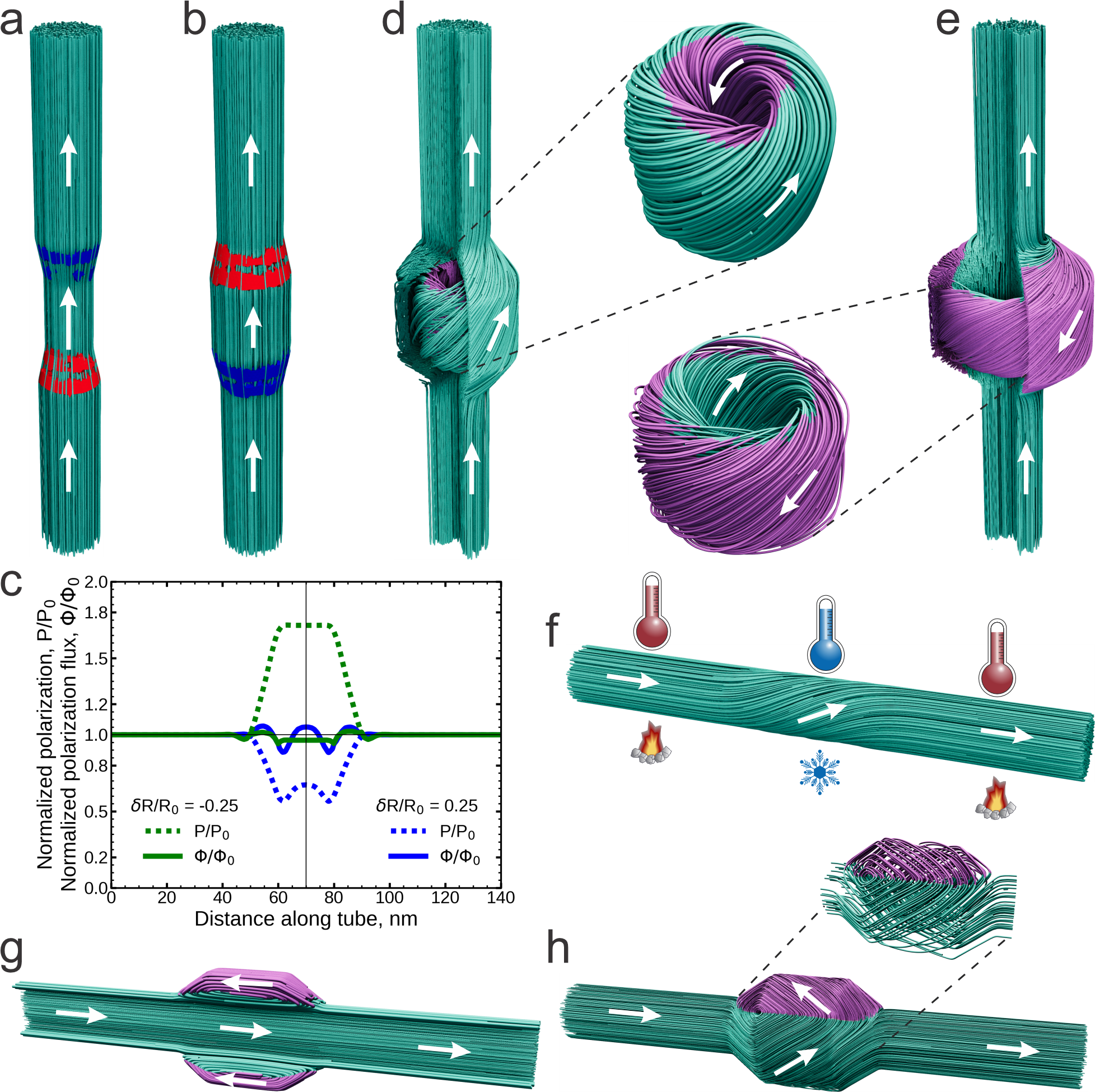}
\caption{\textbf{Modeling of Bernoulli effect in ferroelectric nanorods}. Panels (a)–-(f) correspond to the nearly isotropic ferroelectric, whereas panels (g) and (h) correspond to the ferroelectric with pronounced pseudo-cubic anisotropy.
(a) Local narrowing of the nanorod enhances polarization within the contracted segment. The polarization streamlines are shown in turquoise, the polarization direction is depicted by the white arrows. The emergence of minor positive and negative bound charges in the transition regions, where polarization changes, is highlighted in red and blue, respectively.
(b) Expansion of the nanorod results in a decrease in polarization in the central segment.
(c) Dependence of polarization, $P$, and polarization flux, $\Phi$, along the nanorod axis for the cases of narrowing and broadening. Their values are normalized on the equilibrium values $P_0$, and  $\Phi_0$ at the initial section of the nanorod. Minor oscillations correspond to the effect of bound charges.
(d) Further expansion of the nanotube forms an oppositely directed polarization domain, highlighted in violet within the central part of the expanded region. The twisting of polarization in this domain imparts a Hopfion structure to it. The polarization flux flowing through the nanotube bypasses this internal Hopfion.
(e) An even greater expansion of the nanotube modifies the domain structure. Now, oppositely directed swirled polarization emerges at the edges of the expanded region, forming an external Hopfion. This structure is pierced by the polarization flux flowing through the nanorod.
(f) Local cooling of the nanorod induces the thermal Bernoulli effect, generating a swirling polarization flux within the cooled region.
(g) The uniaxial crystalline anisotropy aligns the polarization streamlines along the anisotropy axis, resulting in an external cylindrical bubble domain.
(h) The cubic anisotropy of the crystal results in the curl-like textures of the polarization streamplines in the wide segment.
}
\label{Fig3}
\end{figure*}

Our further simulations primarily focus on ferroelectric nanorods, a promising system for investigating the ferroelectric Bernoulli effect.
Panels (a)-(f) of Fig.~\ref{Fig3} show the simulation results of the Bernoulli effect in ferroelectric cylindrical nanorods of radius of $R_0=8$\,nm, made of nearly isotropic ferroelectric material. Importantly, we minimize the GLD functional (\ref{Functional}), including the electrostatic contribution. As a result, the system’s tendency to eliminate bound charges in order to minimize energy and conserve the polarization flux along the tube is a confirmable phenomenon. The central segment, with a length of ${l=30}$\,nm, slightly narrows in panel (a) with $\delta {R/R=-0.25}$ and expands in panel (b) with $\delta {R/R=0.25}$. 

The polarization streamlines remain parallel to the nanorod axis within the segments. However, they become more denser in the narrowed segment and less dense in the expanded segment, indicating an increase and decrease in polarization amplitude, respectively, according to the Bernoulli principle. Panel (c) shows the dependence of the polarization amplitude, $P$, and polarization flux, $\Phi = \pi R^2 P$, along the nanorod axis. Notably, the polarization flux is nearly conserved along the nanorod length in both cases, confirming the relation (\ref{EqFlucConsrev}). Small oscillations of the polarization and its flux are linked to the emergence of small residual positive and negative bound charges with a density of $\rho_b \sim \pm 10^{-3}$\,aC\,nm$^{-3}$ in the terminal transition regions of the segment, as indicated by the red and blue colors in panels (a) and (b), where the polarization gradients are most pronounced. However, their electrostatic effect remains minimal.

Phase separation occurs in the broadened segment of the nanorod when this segment expands beyond a threshold value  $\delta R/R \approx 0.35$.  A domain with counter-directed longitudinal polarization components nucleates inside the segment. However, unlike the cylindrical domain of uniaxial polarization described in the previous section, the counter-flux domain in a nearly isotropic ferroelectric material exhibits a more intricate topological structure, as illustrated in panel (d) for $\delta R/R = 0.85$. The polarization streamlines in this region become swirled, acquiring a "tangled ball of yarn" shape, where they curve around one another in a twisted manner. 
This swirled texture of streamlines, known in topology as a Hopfion~\cite{Arnold2021}, cannot be untangled or reduced to a simpler form without altering the topological linkage of the streamlines. In ferroelectric nanotube with broadening, the polar Hopfion structure~\cite{Lukyanchuk2020}  allows the polarization to maintain a nearly constant amplitude, while the flux through the nanorod remains conserved, bypassing the internal Hopfion region, where the total flux is nearly zero.

As the segment radius increases further, surpassing another critical value of $\delta R/R \approx 1.1$, a new swirled configuration emerges. Twisted polarization components with opposite orientations develop at the edges of the expanded region, leading to the formation of an external Hopfion. Similar to its internal counterpart, the external Hopfion represents a localized, entangled knot of space-looped twisted polarization streamlines. It is situated at the periphery of the expanded segment, while the polarization flux through the nanorod remains conserved, piercing the external Hopfion from within. The structure of the fully developed external Hopfion at $\delta R/R = 1.5$ is shown in panel (e).

Another manifestation of the Bernoulli effect arises when a localized region of the nanorod, instead of expanding or contracting, is subjected to thermal influence. For instance, upon cooling the central part of the nanorod, the interplay of flux conservation with the ferroelectric tendency to keep polarization magnitude near equilibrium value, induces a temperature-driven helical twisting of the polarization. Such thermoelastic torsion effect is illustrated in panel (f) for the case where the nanorod segment is cooled below -100$^\circ$C. In this helical configuration, the polarization vector acquires transversal components, allowing it to maintain an amplitude close to the corresponding equilibrium value at lower temperature, while ensuring the constancy of the flux. The emergence of polarization swirling along the axis of a ferroelectric nanotube is a fundamentally new phenomenon with no analogue in fluid dynamics. In the latter, vorticity is constrained by Helmholtz’s theorem, which ensures its conservation during motion~\cite{Landau6,Batchelor2000}. In ferroelectrics, however, no such integral of motion exists, allowing the spontaneous formation of polarization twisting along streamlines.

Now, we examine how the anisotropy of the polar material influences the obtained results. We begin with the case of strong uniaxial anisotropy, which is analogous to the situation qualitatively described in the previous section. Panel (g) presents simulations of the Bernoulli effect in a ferroelectric nanorod made from a uniaxial ferroelectric material. The axis of uniaxial anisotropy is aligned with the axis of the nanorod. The nanorod features a wide segment with the same geometric parameters as those in panel (d). Notably, unlike the internal cylindrical domain bubble bypassed by the principal polarization flux, as discussed in the previous section (see Fig. \ref{Fig2}d), an oppositely oriented polarization region forms at the segment’s periphery, with the polarization flux piercing through the center of the bubble domain. In comparison with the isotropic case, strong uniaxial anisotropy unwinds the external Hopfion, leading to the formation of cylindrical bubble domains where the polarization is nearly entirely aligned with the nanorod axis.

To investigate the influence of crystalline anisotropy in a ferroelectric material with cubic lattice structure, we analyze the behavior of the polarization flux in cylindrical nanorods made of the exemplary barium titanate, BaTiO$_3$, with cubic symmetry in paraelectric phase. We assume that one of the symmetry axes is aligned with the nanorod’s orientation and account for elasticity effects by considering the full GLD functional (\ref{Functional}) in our simulations. For small variations in segment size, the polarization streamlines remain aligned with the nanorod axis, similar to the isotropic case shown in panels (a) and (b). However, as the segment widens, phase separation occurs within the segment, as shown in panel (h). On the outside, this manifests as the emergence of 90$^\circ$-oriented non-equal polarization domains within the segment. Further detailed analysis reveals that these domains are formed by cycloid-like curls of the polarization flux streamlines, as shown in the inset. Notably, the stretching of the nanorod along the axis imparts uniaxial anisotropy to the nanorod, which can lead to the alignment of the lines along the nanorod axis, resulting in the situation shown in panel (g). However, our simulations indicate that such alignment can occur only under unrealistically high strains, above~5\%.

\blue{
While the present simulations consider either nearly isotropic or uniaxial ferroelectrics, many functional materials, such as rhombohedral BiFeO$_3$, exhibit lower crystallographic symmetry. These symmetries may influence the preferred orientation of domains and local anisotropic energy contributions. However, the topological polarization structures described here originate from global polarization flux conservation and are fundamentally nonlocal. As such, their existence and qualitative features are expected to be robust with respect to symmetry class. Extending our modeling framework to lower-symmetry systems would be valuable for capturing material-specific geometries, while the underlying mechanism is expected to remain valid across different symmetry classes.}


\section{Discussion}

In this work, we introduced the Bernoulli effect in ferroelectrics, demonstrating its analogy with classical fluid dynamics. By formulating a generalized Bernoulli equation for polarization flux, we showed that variations in cross-sectional area lead to corresponding changes in polarization amplitude, akin to the velocity-pressure relationship in fluids. This concept represents the fundamental interplay between topology, electrostatic interactions, and material properties. Extending beyond simplified models, it provides a predictive framework for designing and optimizing ferroelectric nanostructures, bridging the gap between theoretical studies and experimental realizations.
Importantly, we focus on thermodynamically stable polarization configurations that emerge in response to geometric constraints and polarization flux conservation in ferroelectric nanostructures. The domain structures and topological textures described in our approach are not the result of non-equilibrium kinetics such as Kibble-Zurek-type dynamics~\cite{Meier2017}, nor do they arise from structural defects~\cite{Dawber2005}, or from stochastic fluctuations~\cite{Scott2007}. Instead, these states are obtained through variational minimization of the GLD functional and are governed by electrostatic and GL energy balance under flux-conserving constraints. As such, they represent a robust class of equilibrium responses in nanoscale ferroelectrics and are fully consistent with the experimentally observed stabilization of topological structures in modern ferroelectric nanomaterials.

The conservation of polarization flux increases polarization in constricted regions and decreases in expanded regions. A key finding is the emergence of localized topological states in ferroelectrics when the cross-sectional area of a polarization flux surpasses a critical threshold. Analogous to cavitation in fluids, this instability gives rise to topological polarization structures such as vortex domains, polarization bubbles and curls and Hopfions. These structures emerge as the system seeks to minimize depolarization energy, which would otherwise dominate over the bulk GL energy. Our phase-field simulations confirm that the nearly divergence-free polarization condition plays a crucial role in shaping these states.

Our results offer new insights into polarization dynamics and deepen the understanding of topological states in ferroelectrics. Future research should prioritize experimental validation using Piezoresponse Force Microscopy (PFM), Scanning Transmission Electron Microscopy (STEM), electron holography, and X-ray nanodiffraction, while also exploring the effects of strain and flexoelectric coupling. Investigating confined geometries in ferroelectric nanorods and nanowires may reveal additional functional possibilities.

Previous studies~\cite{Lich2022,Liu2017,Lahoche2008,Kondovych2023}  have demonstrated that ferroelectric nanorods and nanowires exhibit a diverse range of topological polarization states, including vortex and helical chiral phases, with surface tension effects playing a crucial role in the stability and formation of these phases. Our findings further extend these studies by revealing how the Bernoulli effect in ferroelectric nanorods and nanowires with variable cross-sections induces polarization flux conservation, leading to the emergence of new topological polarization structures. Moreover, our results suggest the possibility of a novel thermoelastic torsion effect in ferroelectric nanotubes and nanowires, where temperature variations could induce helical polarization twisting. These effects offer an innovative mechanism for polarization control, potentially enabling the development of advanced functional nanoelectromechanical systems. 
Note that while our simulations are carried out at the tens-of-nanometers scale for computational convenience, the underlying thermodynamic mechanisms are scale-invariant and readily extend to larger structures. The model naturally scales to broader size ranges, making the predicted phenomena applicable to a wide variety of experimentally relevant geometries.

Our findings extend the application of the ferroelectric Bernoulli effect to soft ferroelectrics, including recently discovered ferroelectric nematic liquid crystals~\cite{Nishikawa2017,Chen2020,Kumari2024}. In these materials, spontaneous polarization responds freely to sample geometry and electrostatic constraints, making the ferroelectric Bernoulli effect a key mechanism for controlling polarization structures. Similar to how variations in cross-sectional area redistribute polarization in solid ferroelectrics, variable thickness of the cell or local changes in dielectric permittivity in ferroelectric nematics can induce polarization twisting and the formation of stable topological structures. These effects open new possibilities for dynamic polarization control in soft ferroelectric materials, with promising applications in tunable electro-optic devices, adaptive optical elements, and sensors based on controllable topological states. 
\blue{It is essential to note that the present study is limited to the static regime, where the polarization field minimizes a static free energy functional. Thus, dynamical aspects, such as viscous relaxation in ferroelectric liquid crystals or kinetic effects in solid ferroelectrics, are not considered here. The applicability of the flux-conservation concept to dynamical regimes would require additional analysis incorporating viscoelastic or hydrodynamic couplings.
}

\blue{In our simulations, we considered freestanding nanorods without surface screening, to capture intrinsic polarization flux behavior driven by depolarization energy. In experiments, surface charge compensation via electrodes, adsorbates, or ambient screening can partially relax the $\mathrm{div}\,\mathbf{P} \approx 0$ condition, altering the geometry or threshold of topological states. Mechanical clamping or substrate embedding may also impose additional anisotropies. While such boundary effects may shift quantitative details, the underlying approximate flux-conserving mechanism and associated topological behavior are expected to remain valid within a substantial range of conditions.}
\blue{The mechanism proposed in this work, based on polarization flux conservation, provides a concrete physical principle that may guide future experimental exploration. We hope that our results will stimulate further efforts using advanced techniques such as STEM, PFM, and coherent X-ray imaging to detect and characterize these topological states in confined ferroelectric geometries.}

\section{Conclusion}

This work establishes the Bernoulli effect as a fundamental principle governing polarization flux in ferroelectrics, revealing its direct impact on polarization distribution in constrained geometries. We demonstrated that changes in cross-sectional area lead to polarization redistribution, analogous to velocity-pressure relationships in fluid dynamics, and that this mechanism drives the formation of diverse topological polarization states, including vortex domains, polarization bubbles and curls, and Hopfions. Moreover, we extended this concept to ferroelectric nematic liquid crystals, where polarization twisting emerges as a natural consequence of flux conservation, enabling the control of mesoscale topological structures. 
\blue{
The mechanism of polarization flux conservation described in this work offers a predictive framework for the formation of topological polarization textures in confined ferroelectric structures. The results presented here may serve as a basis for motivating future experimental studies aimed at identifying and characterizing these textures using advanced imaging techniques.
}
\blue{Although our analysis is restricted to equilibrium states, its extension to non-equilibrium conditions (e.g., field-driven switching) represents a natural next step toward exploring the kinetic stability and switching pathways of topological polarization structures.
}
These insights provide a foundation for engineering functional ferroelectric nanostructures and soft-matter devices, with potential applications in nanoelectronics, photonics, and reconfigurable electro-optic systems.

\section{Phase-field modeling}
\label{Appendix}

\subsection{Functional.} 
To model the polarization field in ferroelectric nanorods, we employed the phase-field method. The governing equations are derived from the minimization of the GLD free-energy functional (\ref{Functional}) for a pseudocubic ferroelectric material, incorporating both elastic and electrostatic contributions. The corresponding energy terms  for pseudo-cubic ferroelectric material are  given by:

a) Ginzburg-Landau energy,
\begin{equation}
\begin{split}
F_{GL} &= \alpha_{1}(T-T_c)(P_{1}^{2}+P_{2}^{2}+P_{3}^{2})+\alpha_{11}(P_{1}^{4}+P_{2}^{4}+P_{3}^{4}) \\
&+\alpha_{12}(P_{1}^{2}P_{2}^{2} + P_{1}^{2}P_{3}^{2}+P_{2}^{2}P_{3}^{2})+\alpha_{111}(P_{1}^{6}+P_{2}^{6}+P_{3}^{6}) \\
&+\alpha_{112}[P_{1}^{4}(P_{2}^{2}+P_{3}^{2})+P_{2}^{4}(P_{1}^{2}+P_{3}^{2})+P_{3}^{4}(P_{1}^{2}+P_{2}^{2})]+\alpha_{123}P_{1}^{2}P_{2}^{2}P_{3}^{2}.
\label{GL}
\end{split}
\end{equation}

b) Polarization gradient energy,
\begin{equation}
 F_{grad} = \frac{1}{2}G_{ijkl}(\partial_{i}P_{j})( \partial_{k}P_{l}).
\label{Grad}
\end{equation} 

c) Electrostatic term which includes the coupling of the polarization with the electric field, described by the electrostatic potential $\mathbf{E}=-\nabla \varphi$ and the proper electrostatic energy, 
\begin{equation}
 F_{\varphi} = \left( \partial_{i}\varphi \right)P_{i} - \frac{1}{2}\varepsilon_{0}\varepsilon_{b}{(\nabla\varphi)}^{2}.
\label{Phi}
\end{equation} 

d) Elastic  term  which includes the coupling of the polarization with the elastic strain $u_{ij}$ and the proper elastic energy,  
\begin{equation}
 F_{elast} = - C_{ijkl}Q_{klmn}u_{ij}P_{m}P_{n} + \frac{1}{2}C_{ijkl}u_{ij}u_{kl}. 
\label{Elast}
\end{equation} 
Here the repeated indices $i,j,...$=$1,2,3$ (or $x,y,z$) indicate the summation over these indices.

The electrostatic potential and strain tensor are denoted as $\varphi$ and $u_{ij}$ respectively. The value of the vacuum permittivity $\varepsilon_0$ is $8.85 \times 10^{-12} CV^{-1}m^{-1}$. The numerical values of the material-specific Ginzburg-Landau expansion coefficients $\alpha_{ijk}$, gradient energy coefficients $G_{ijkl}$, elastic stiffness tensor $C_{ijkl}$, tensor of electrostrictive coefficients $Q_{ijkl}$ and background dielectric constants $\varepsilon_b$ are given below.

\subsection{Material coefficients.} 
For the simulation of model isotropic ferroelectric materials, we selected material coefficients with numerical values comparable to those of the typical ferroelectric material PbTiO$_3$, while ensuring that the functional~(\ref{Functional}) retains its rotational invariance~\cite{Kondovych2023}. Specifically, we used $\alpha_1 = 3.8\times 10^5 \, C^{-2}m^2N$, $T_c=479^{\circ}$C,  $\alpha_{11} = 0.41 \times 10^9 \, C^{-4}m^6N$, and $\alpha_{12} = 2\alpha_{11} = 0.82 \times 10^9 \, C^{-4}m^6N$, and neglected the sixth-order GL coefficients. The elastic contribution was incorporated into the coefficient $\alpha_{12} = 2\alpha_{11}$ following the renormalization procedure, described in Section~\ref{GLapproach}. 
The gradient energy coefficients 
$G_{1111}$ = $2.77\times 10^{-10} C^{-2}m^{4}N$,
$G_{1122}$ = 0,
$G_{1212}$ = $1.38\times 10^{-10} C^{-2}m^{4}N$ were selected from~\cite{Li2002,Wang2004}. 
The background dielectric constant is $\varepsilon_b\approx 10$~\cite{Mokry2016}. 
In temperature-dependent simulations, the temperature distribution along the nanorod axis $x$ was modeled as  
$
T(x) = 25^\circ\mathrm{C} - \exp\left(-\frac{x^2}{2\gamma^2}\right)\times 125^\circ\mathrm{C} 
$
with $\gamma = 16$\,nm. This results in a temperature of $T = -100^\circ\mathrm{C}$ at the center of the nanorod ($x=0$) and room temperature ($T = 25^\circ\mathrm{C}$) at the ends.

To induce uniaxial anisotropy, we used the same simulation setup as described above, but modified the quadratic terms in the functional (\ref{GL}) as $\alpha_{1}(T-T_c)P_{1}^{2}+a_2 P_{2}^{2}+a_3 P_{3}^{2}$, with positive coefficients $a_2$=$a_3$=$0.17 \times 10^9 \, \textrm{C}^{-2}\textrm{m}^2\textrm{N}$.

The coefficients of the full anisotropic GLD functional~(\ref{Functional}) for BaTiO$_3$ ferroelectric were taken from~\cite{Rabe2007, HlinkaBTO2006} as follows: 
$\alpha_{1}$ = $3.34\times 10^{5} C^{-2}m^{2}N, T_c=108^{\circ}C$,
$\alpha_{11}$ = $1.7\times 10^{8} C^{-4}m^{6}N$,
$\alpha_{12}$ = $-3.44\times 10^{8} C^{-4}m^{6}N$,
$\alpha_{111}$ = $8.004\times 10^{9} C^{-6}m^{10}N$,
$\alpha_{112}$ = $4.47\times 10^{9} C^{-6}m^{10}N$,
$\alpha_{123}$ = $4.91\times 10^{9} C^{-6}m^{10}N$. 
The electrostrictive tensor coefficients are given by: 
$Q_{1111}$ = $11.04\times 10^{-2} C^{-2}m^{4}$,
$Q_{1122}$ = $ - 4.52\times 10^{-2} C^{-2}m^{4}$,
$Q_{1212}$ = $1.445\times 10^{-2} C^{-2}m^{4}$.
The components of the elastic stiffness tensor are: 
$C_{1111}$ = $27.5\times 10^{10} m^{-2}N$,
$C_{1122}$ = $17.9\times 10^{10} m^{-2}N$,
$C_{1212}$ = $5.43\times 10^{10} m^{-2}N$.
The gradient energy coefficients are:
$G_{1111}$ = $51\times 10^{-11} C^{-2}m^{4}N$,
$G_{1122}$ = $ - 2\times 10^{-11} C^{-2}m^{4}N$,
$G_{1212}$ = $2\times 10^{-11} C^{-2}m^{4}N$.
The background dielectric constant is $\varepsilon_b \approx 7.35$.
\blue{To ensure physical relevance, the chosen nanorod radius lies between the coherence length, typically on the order of 1-2\,nm, and the internal screening depth, which may exceed several tens of nanometers in ferroelectric perovskites with semiconducting charges}.

\blue{Note that material-to-material differences of the material parameters in the functional ferroelectric oxides are relatively small. As such, the qualitative features of the polarization textures are expected to remain stable across realistic material systems. Specifically, the gradient coefficients $ G_{ijkl}$ set the coherence length of the polarization field. Since the characteristic size of the nanorods and the emerging topological structures is much larger than this coherence length, the observed effects are stable with respect to moderate variations of these parameters.  Therefore, the developed modeling methodology remains adaptable for simulations in materials with similar material parameters, should experimental conditions require it.}

\subsection{Phase-field modeling.} 
The non-linear differential relaxation equation is used to find the minima of the free-energy (\ref{Functional}):

\begin{equation}
   - \gamma\frac{\partial P}{\partial t} = \frac{\delta F}{\delta P}\,\,\,.
    \label{Variation}
\end{equation}
Here $\gamma$ denotes a time-scale parameter set to unity. The non-linear terms are coupled with two linear systems governed by the screened Poisson equation~(\ref{Poisson}) and linear elasticity equations~(\ref{LinearElasticity}) in BTO simulations.
The distribution of the electrostatic potential $\varphi$ and the elastic strains $u_{ij}$ is found from the respective electrostatic and elastic equations that are obtained by the respective variation of the GLD functional~(\ref{Functional}) over $\varphi$ and $u_{ij}$:
\begin{equation}
    \varepsilon_{0}\varepsilon_{b}\nabla^{2}\varphi = \partial_{i}P_{i}\,\,\, ,
   \label{Poisson}
\end{equation}
\begin{equation}
   C_{ijkl}\partial_{j}\left( u_{kl} - Q_{klmn}P_{m}P_{n} \right) = 0.
    \label{LinearElasticity}
\end{equation}

Phase-field modeling was performed using the FEniCS computational framework~\cite{LoggMardalEtAl2012a}. Three-dimensional cylindrical domains were discretized using unstructured tetrahedral finite-element meshes generated with the \textit{gmsh}~\cite{Geuzaine2009}. Solutions for \textbf{P}, $\varphi$, and $u_{ij}$ were computed within piecewise-linear polynomial function spaces. Both polarization \textbf{P} and potential $\varphi$ were subject to periodic boundary constraints along the \textit{z}-axis.

The temporal derivative in (\ref{Variation}) was discretized via a BDF2 adaptive time-stepping scheme~\cite{Janelli2006}. Initial polarization conditions involved random perturbations of polarization vector components within $[-10^{-6}, 10^{-6}]$ Cm$^{-2}$. Non-linear systems from (\ref{Variation}) were addressed using Newton-Raphson iterations with line search. Linear systems 
of~(\ref{Poisson}) and~(\ref{LinearElasticity}) at each iteration, were solved using the restarted generalized minimal residual (GMRES) method~\cite{petsc-user-ref,petsc-web-page}.

\vspace{12pt}

\textbf{Funding} This research was funded by European Union HORIZON actions:  MSCA-RISE-MELON (project number 872631) and  MSCA-SE-H-GREEN (project number 101130520).
A.R. acknowledges the Slovenian Research Agency support (P1-0125).

\textbf{Data availability} Data underlying the results presented in this paper are not publicly available at this time but may be obtained from the authors upon reasonable request.

\textbf{Acknowledgments} This work was granted access to HPC resources of “Plateforme MatriCS” within University of Picardie Jules Verne. “Plateforme MatriCS” is co-financed by the European Union with the European Regional Development Fund (FEDER) and the Hauts-De-France Regional Council among others. The authors gratefully acknowledge Prof. Oleg Lavrentovich for his insightful suggestion regarding the applicability of the ferroelectric Bernoulli effect in ferroelectric nematic liquid crystals.

\printbibliography

\end{document}